\documentclass[aps,plb,superscriptaddress,twoside,showpacs,
nofootinbib,10pt,floatfix,twocolumn,showpacs]{revtex4-1}
\usepackage{graphicx}
\usepackage{mathrsfs}
\usepackage{amsmath,bbm}
\usepackage{CJK}
\usepackage{tikz}
\usepackage[colorlinks,hyperindex,citecolor=black,
	linkcolor=black,urlcolor=black]{hyperref}

\newcommand\<{\langle}
\renewcommand\>{\rangle}
\renewcommand\d{\partial}
\newcommand\tr{\mathop{\mathrm{Tr}}}

\begin{document}

\begin{CJK*}{UTF8}{gbsn}
\title{Skyrme model study of proton and neutron properties in a strong magnetic field}
\author{Bing-Ran He (何秉然)}
\email[E-mail: ]{hebingran@njnu.edu.cn}
\affiliation{
Department of Physics, Nanjing Normal University, Nanjing 210023, P.R. China
}

\date{\today}

\begin{abstract}
\end{abstract}
\begin{abstract}
The proton and neutron properties in a uniform magnetic field are investigated. 
The Gell-Mann-Nishijima formula is shown to be satisfied for baryon states. It is found that with increasing magnetic field strength, the proton mass first decreases and then increases, while the neutron mass always increases. 
The ratio between magnetic moment of proton and neutron increases with the increase of the magnetic field strength. 
With increasing magnetic field strength, the size of proton first increases and then decreases, while the size of neutron always decreases. The present analyse implies that in the core part of the magnetar, the equation of state depend on the magnetic field, which modifies the mass limit of the magnetar. 
\end{abstract}



\maketitle
\end{CJK*}
\section{Introduction}
Recently, experiments have observed that there exists a strong magnetic field when baryon collide with each other, and astrophysics have observed that the strong magnetic field exists in magnetars~\cite{Kharzeev:2012ph,Miransky:2015ava,Skokov:2009qp}. 
The baryon states have electric charge distribution, thus the interaction between baryons and magnetic fields modifies the properties of baryons. 
The Skyrme model~\cite{Skyrme:1962vh}, which identifies the soliton solution from mesons theories as the baryon, has been widely accepted, and also have lots of applications to hadron physics, astrophysics and also condensed matter physics. 
The study of Skyrmion in a uniform magnetic field shows that, in the leading order of large $N_C$, i.e., $\mathcal{O}(N_C)$, the mass and shape of Skyrmion depend on the strength of magnetic field~\cite{He:2015zca}. 

In this letter, the $\mathcal{O}(N_C^{-1})$ effects are introduced in the semi-classical quantization approach~\cite{Adkins:1983ya}, and then the physical baryon states, i.e, proton and neutron, in a uniform magnetic field are studied. 
The Gell-Mann-Nishijima formula for baryon states are shown to be satisfied. 
The semi-classical quantization of Skyrmion introduces time dependence to $(eB)$ terms of the model. Because the wave functions for baryon states are different, the magnetic response of baryon states are different.  
It is found that with the increase of the magnetic field strength, the effective proton mass first decreases and then increases, consequently, the proton size first increases and then decreases. On the other hand, the effective neutron mass always increases, and  consequently, the neutron size always decreases. 
Furthermore, the ratio between magnetic moment of proton and neutron increases with the increase of the magnetic field strength. 
Finally, since both the mass and size of proton and neutron depend on the strength of the magnetic field, the equation of state for magnetar is modified. 

\section{The model}

The action of the model contains two parts:
\begin{equation}
\Gamma=\int d^4 x \mathscr{L}+\Gamma_{\rm WZW}\,,
\label{total_action}
\end{equation}
where $\mathscr{L}$ is expressed as
\begin{eqnarray}
\mathscr{L}&=&\frac{f_\pi^2}{16}\tr(D_\mu U ^\dag D^\mu U) + \frac{1}{32g^2}\tr([U ^\dag D_\mu U, U^\dag D_\nu U]^2)\nonumber\\
&&+\frac{m_\pi^2 f_\pi^2}{16}\tr(U+U^\dag-2)
\,.\label{lagrangian}
\end{eqnarray}
Here $f_\pi$ is the pion decay constant, $m_\pi$ is the pion mass, and $g$ is a dimensionless coupling constant. 
The covariant derivative for $U$ is expressed as $D_\mu U = \partial_\mu U - i \mathcal{L}_\mu U + i U \mathcal{R}_\mu$,  
where $\mathcal{L}$ and $\mathcal{R}$ are the external fields expressed as  $\mathcal{L}_\mu=\mathcal{R}_\mu= e Q_{\rm B} {\mathcal{V}_B}_\mu + e Q_{\rm E} H_\mu$ for the present purpose. Here  
$e$ is the unit electric charge, $Q_{\rm B}=\frac{1}{3}\mathbbm{1}$ is the baryon number charge matrix, $Q_{\rm E}=\frac{1}{6}\mathbbm{1}+\frac{1}{2}\tau_3$ is the electric charge matrix,  $\mathbbm{1}$ is the rank $2$ unit matrix, $\tau_3$ is the third Pauli matrix, and ${\mathcal{V}_B}_\mu$ is the external gauge field of the $U(1)_\mathcal{V}$ baryon number. 
In the symmetric gauge, the magnetic field $H_\mu$ is expressed as $
H_\mu  = - \frac{1}{2} B  y \eta_\mu^{\;\;1} + \frac{1}{2} B  x \eta_\mu^{\;\;2}
$, 
where $\eta$ is the geometry with ${\rm diag}(+1,-1,-1,-1)$. 

The Wess-Zumino-Witten (WZW) action $\Gamma_{\rm WZW}\equiv\int d^4x \mathscr{L}_{\rm WZW}$, represents the chiral anomaly effects, which is given in Refs.~\cite{Wess:1971yu,Witten:1983tw}.

\section{The semi-classical quantization}
Following Ref.~\cite{Holzwarth:1985rb}, the $x$, $y$, and $z$ in the elliptic coordinate system are expressed as 
\begin{eqnarray}
x&=&c_\rho r \sin (\theta )\cos (\varphi ) \,,\nonumber\\
y&=&c_\rho r \sin (\theta )\sin (\varphi ) \,,\nonumber\\
z&=&c_z r \cos (\theta )\,,\label{xyzansatzs}
\end{eqnarray}
where $c_\rho$ and $c_z$ are positive dimensionless parameters, 
$r\equiv \sqrt{\frac{x^2}{c_\rho^2} + \frac{y^2}{c_\rho^2} + \frac{z^2}{c_z^2}}$, and $\theta$ and $\varphi$ are polar angles with $\theta\in[0,\pi]$ and $\varphi\in[0,2\pi]$. 
The $U$ is decomposed in the Cartesian coordinate system as 
\begin{eqnarray}
U=\cos(F(r))\mathbbm{1} + \frac{i \sin(F(r))}{r}\Big(  \frac{\tau_1}{c_\rho}x + \frac{\tau_2}{c_\rho}y +  \frac{\tau_3}{c_z}z\Big)\,.\label{uansatzs}
\end{eqnarray}

The ansatz equations \eqref{xyzansatzs} and \eqref{uansatzs} are the solution of the statical case of Skyrmion, the physical baryon states are obtained by semi-classical quantization of the Skyrme model. 
The quantization of single Skyrme model is proposed by Ref.~\cite{Adkins:1983ya}, and then the discussion is extended to many baryon states~\cite{Braaten:1988cc,Krusch:2002by}. 
Following~\cite{Braaten:1988cc}, the time dependence of $U$ is expressed as 
\begin{eqnarray}
\hat{U} &=& A(U (R))A^\dag \,, 
\label{ULR_t}
\end{eqnarray}
where $A$ is the rotation matrix of isospin space,  
and $R$ is the rotation matrix of spatial space in $x-y$ plane for the present purpose.  
The rotation matrix $A$ and $R$ are expressed as
\begin{eqnarray}
A^{-1} \dot A = \frac{i}{2} \omega_a \tau_a \,, \hspace{2mm}
(R^{-1} \dot R)_{ij} = -\epsilon_{ij3} \Omega_3 \,,
\end{eqnarray}
where $a=1,2,3$ and $i,j=1,2$. 

Insert \eqref{ULR_t} in action \eqref{total_action} we obtain $\hat\Gamma = \int d^4x (\hat {\mathscr{L}} + \hat {\mathscr{L}}_{\rm WZW}) = \int d^4x \hat {\mathscr{L}}_{\rm total}$. The canonical conjugate momenta of the isospin and spin are obtained by taking a functional derivative of the action with $\omega_a$ and $\Omega_3$, respectively, as 
\begin{eqnarray}
I_a
=\left.\frac{\d \hat {\mathscr{L}}_{\rm total}
}{\d\omega_a}\right|_{{\mathcal{V}_B}_\mu\to0}
\,,\quad\quad
J_3
=\left.\frac{\d \hat {\mathscr{L}}_{\rm total}
}{\d\Omega_3}\right|_{{\mathcal{V}_B}_\mu\to0}
\,.
\end{eqnarray}

\section{The Gell-Mann-Nishijima formula}
The baryon number current of the model is obtained by taking a functional derivative of the WZW term 
with ${\mathcal{V}_B}_\mu$, i.e., $j_B^\mu=\frac{\partial\hat{\mathscr{L}}_{\rm WZW}}{\partial(e{\mathcal{V}_B}_\mu)}|_{{\mathcal{V}_B}_\mu\to0}$. 
The baryon number $N_B$ is obtained as 
\begin{eqnarray}
N_B&=&\int dV j_B^0\nonumber\\
&=&\frac{\sin (2 F) \left(e B c_{\rho }^2 r^2 D_{33} +6\right)-12 F}{12 \pi }\Big|^{F(\infty)=0}_{F(0)=\pi} \nonumber\\
&=&1\,,
\end{eqnarray}
where $dV=c_\rho^2 c_z r^2 \sin (\theta ) dr d\theta d\varphi$ and $D_{33}=\frac{1}{2}\tr[A^\dag\tau_3 A\tau_3]$.  
In the present calculation, the boundary conditions 
 $F(0)=\pi$ and $F(\infty)=0$ are imposed. 

Considering the external fields have a fluctuation as $\mathcal{L}_\mu=\mathcal{R}_\mu= e Q_{\rm B} {\mathcal{V}_B}_\mu + e Q_{\rm E} H_\mu - \delta(\mathcal{V}^a_\mu)\frac{\tau^a}{2}$, 
the corresponding iso-vector current is obtained as 
$j_\mathcal{V}^{a,\mu}=\frac{\partial(\hat{\mathscr{L}}_{\rm total})}{\partial(\delta(\mathcal{V}^a_\mu))}|_{{\mathcal{V}_B}_\mu\to0,\delta(\mathcal{V}^a_\mu)\to0}$. 
The conserved charge corresponding to the third component of $SU(2)$ iso-vector current is obtained as 
\begin{eqnarray}
N_{\mathcal{V}^{3,0}}&=&\int dV j_\mathcal{V}^{3,0}\nonumber\\
&=& - \frac{\sin (2 F) \left(e B c_{\rho }^2 r^2 D_{33} \right)}{24 \pi }\Big|^{F(\infty)=0}_{F(0)=\pi} + I_3 \nonumber\\
&=& I_3\,.
\end{eqnarray}

The Gell-Mann-Nishijima formula for electric charge of baryon is given as
\begin{equation}
N_E=\int dV \left(\frac{j_B^0}{2}+j_\mathcal{V}^{3,0}\right)=
\frac{N_B}{2}+I_3 \,. 
\label{gmn_formular}
\end{equation}
Here the electric charges of baryon states are shown to be always conserved in a uniform magnetic field, which is consistent with the fact that both $U(1)_{\mathcal{V}}$ and the third component of $SU(2)_{\rm isospin}$ symmetries are conserved, respectively. 

\section{Numerical results }
The parameters $c_\rho$ and $c_z$ in Eqs.~(\ref{xyzansatzs}, \ref{uansatzs}) have two effects: deform the ansatz and scale the volume. Since the scale effect of $c_\rho$ and $c_z$ can be absorbed by performing the scale transform of $r$, with no lose of generality, the restriction between $c_\rho$ and $c_z$ is imposed as $c_\rho\equiv 1/\sqrt{c_z}$.

In $N_B=1$ sector, following Ref.~\cite{Adkins:1983ya}, after properly choosing the Skyrme units, a standard set of parameters is considered: 
$m_\pi=138$ [MeV], $f_\pi=108$ [MeV], and $g=4.84$. 

The equation of motion for proton and neutron are obtained from $\<{\Psi}|\hat\Gamma|{\Psi}\>$ at $\mathcal{O}(N_C)$ order, respectively. 
Here $|{\Psi}\>$ expresses the wave functions for $|p\uparrow\>$, $|p\downarrow\>$, $|n\uparrow\>$ and $|n\downarrow\>$ which are given in Ref.~\cite{Adkins:1983ya}. The $N_C$ counting for the parameters of the present model are $f_\pi\sim\mathcal{O}(N_C^{1/2})$, $g\sim\mathcal{O}(N_C^{-1/2})$, $m_\pi\sim\mathcal{O}(N_C^{0})$, $eB\sim\mathcal{O}(N_C^{0})$, $\omega_a\sim\mathcal{O}(N_C^{-1})$ and $\Omega_3\sim\mathcal{O}(N_C^{-1})$.

The Hamiltonian up to $\mathcal{O}(N_C^{-1})$ is obtained as 
\begin{eqnarray}
{\mathcal{H}}= \sum_{a=1,2,3} \left(\omega_a I_a\right) + \Omega_3 J_3 - \left.\hat{\mathscr{L}}_{\rm total}\right|_{{\mathcal{V}_B}_\mu\to0}\,. 
\end{eqnarray}

The nucleon mass and the nucleon magnetic moment are defined as $M_{\Psi}\equiv \<{\Psi}|\int dV \mathcal{H}|{\Psi}\>$ and $\mu_{\Psi}\equiv-\frac{\d M_{\Psi}}{\d (e B)}$, respectively. 
It is easy to check that  $M_{p}\equiv M_{p\uparrow}=M_{p\downarrow}$,  $M_{n}\equiv M_{n\uparrow}=M_{n\downarrow}$,  $\mu_p\equiv\mu_{p\uparrow}=-\mu_{p\downarrow}$ and $\mu_n\equiv\mu_{n\uparrow}=-\mu_{n\downarrow}$.

\begin{figure}[htb]
\centering
\includegraphics[scale=0.42]{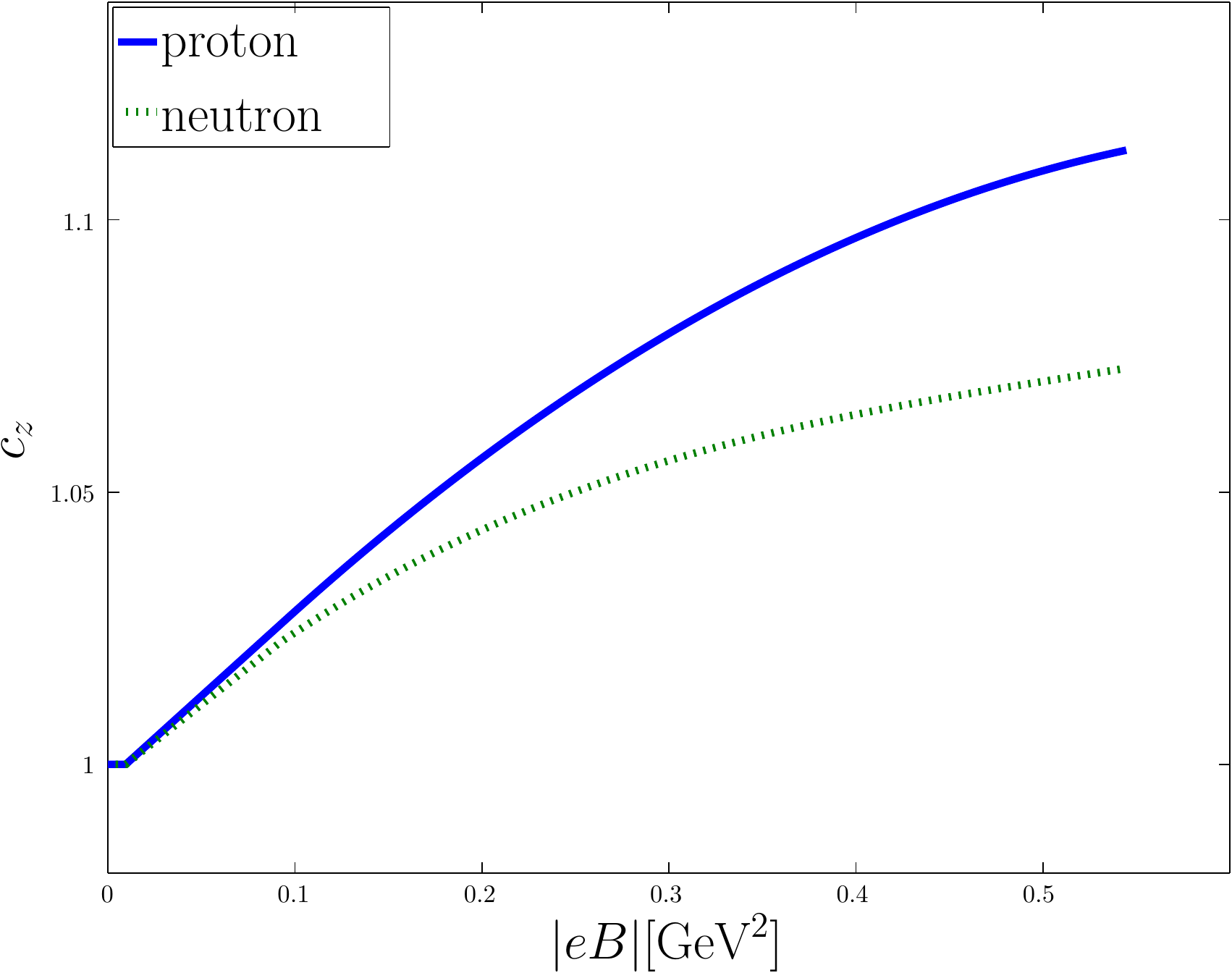} 
\caption{$|e B|$ dependence of $c_z$ for proton and neutron.
}
\label{fig:bmag_cz}
\end{figure}

The parameter ${c}_z$ is fixed to minimize the proton and neutron mass for a given $|e B|$, respectively. 
The $|e B|$ dependence of ${c}_z$ for proton and neutron are shown in Fig.~\ref{fig:bmag_cz}. 
Fig.~\ref{fig:bmag_cz} shows that when the strength of the magnetic field increases, $c_z$ increases, which implies that the shape of proton and neutron are twisted.

\begin{figure}[htb]
\centering
\includegraphics[scale=0.42]{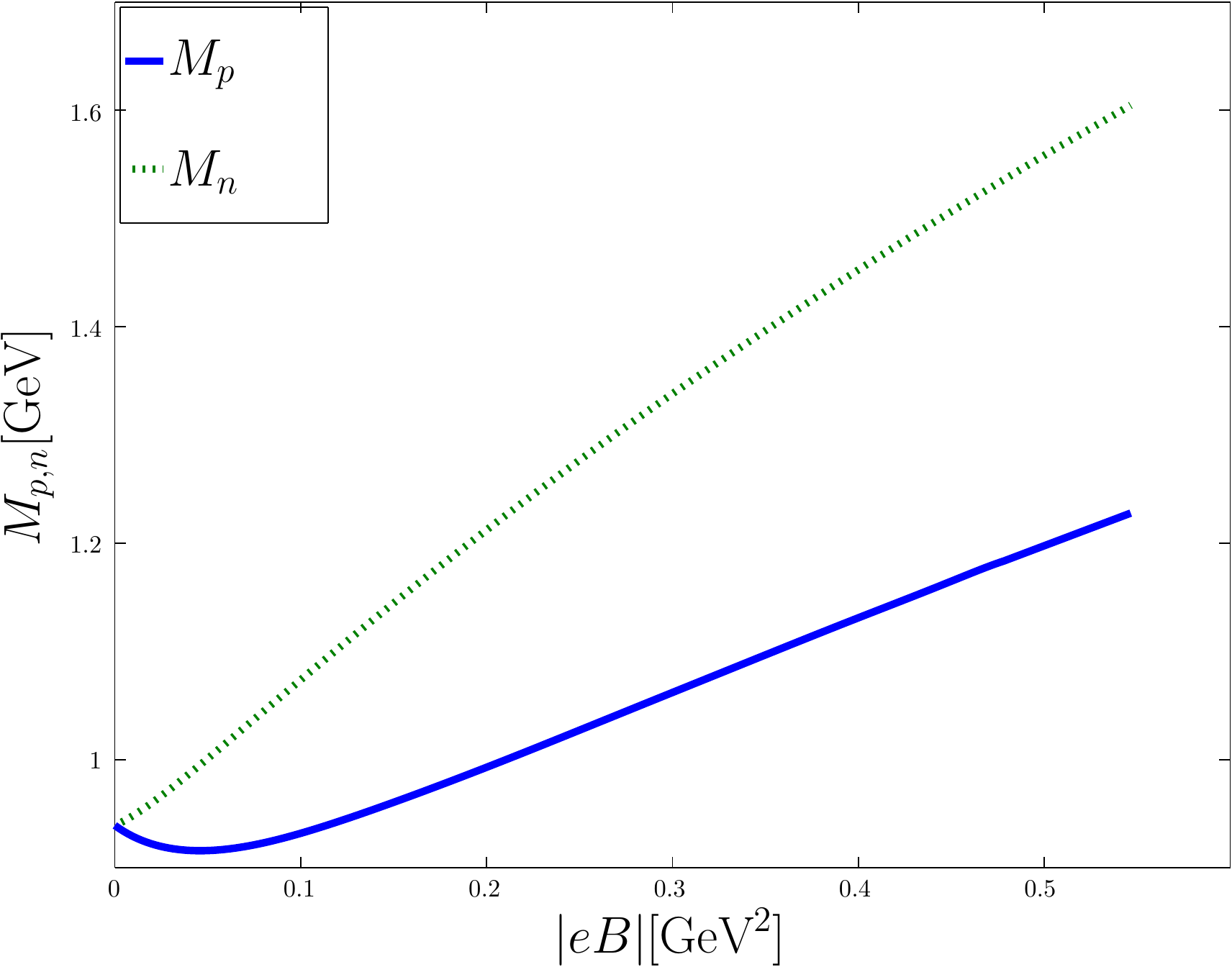} 
\caption{$|e B|$ dependence of $M_{p}$ and $M_{n}$.
}
\label{fig:bmag_mpmn}
\end{figure}

The $|e B|$ dependence of proton and neutron mass are shown in Fig.~\ref{fig:bmag_mpmn}. 
Fig.~\ref{fig:bmag_mpmn} shows that, with increasing magnetic field strength, the proton mass first decreases then increases. This is because of that the Hamiltonian of proton contains linear them of $(eB)$ and higher order terms of $(eB)$, the linear them of $(eB)$ has a different sign with higher order terms of $(eB)$. Therefore, for a weak $|eB|$, the linear term of $(eB)$ takes a dominant role which causes the proton mass to decrease; for a strong $|eB|$, with the increase of $|eB|$, the dominant role is shifted to higher order terms of $(eB)$, which causes the proton mass to increase. 
Fig.~\ref{fig:bmag_mpmn} also shows that, the neutron mass always increases when the magnetic field strength $|eB|$ increases, since in the Hamiltonian of neutron, the linear term of $(eB)$ has a same sign with higher order terms of $(eB)$, which causes the neutron mass to increase. 

\begin{figure}[htb]
\centering
\includegraphics[scale=0.42]{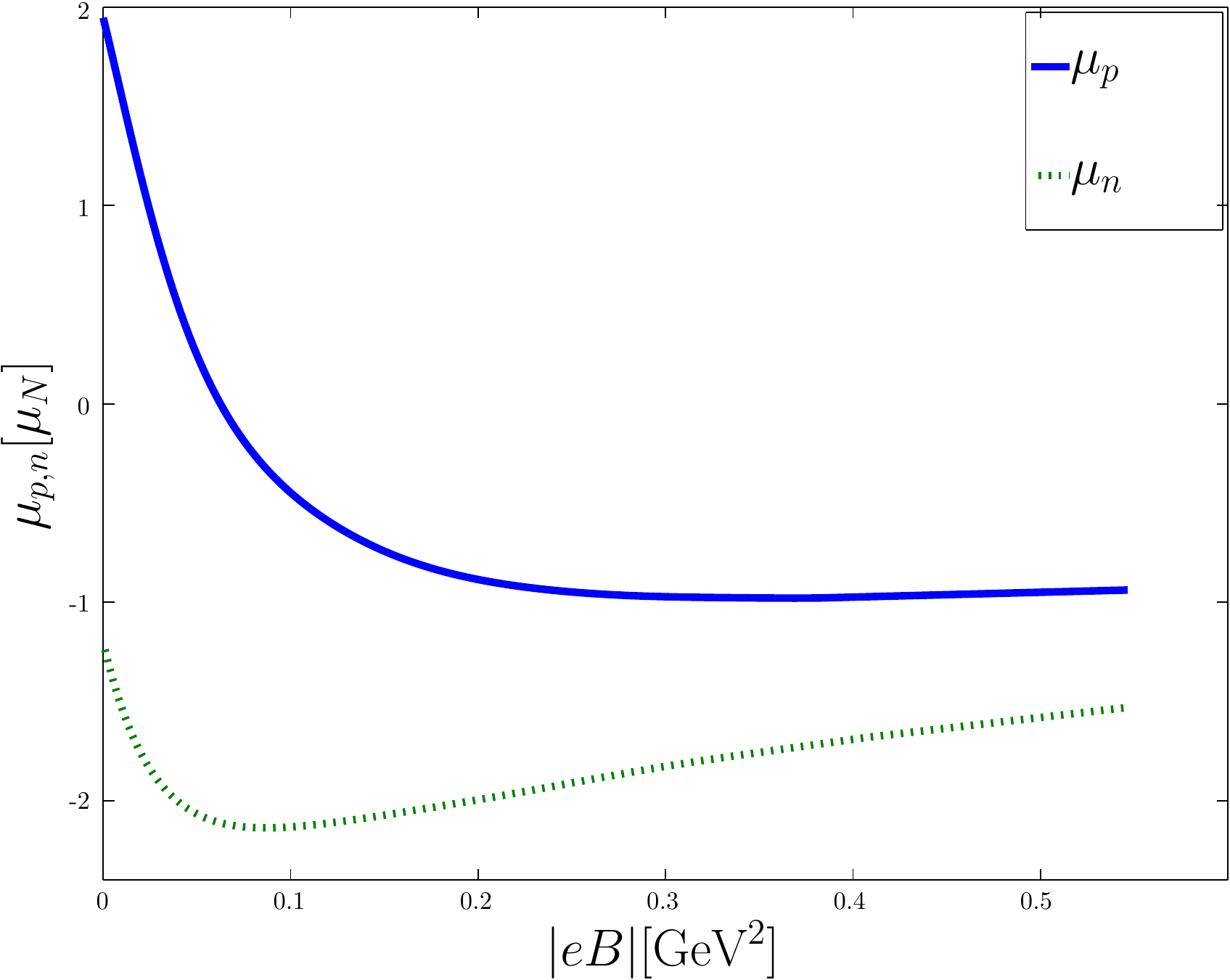} 
\caption{$|e B|$ dependence of $\mu_{p}$ and $\mu_{n}$.
}
\label{fig:bmag_mupmun}
\end{figure}

The $|e B|$ dependence of $\mu_{p}$ and $\mu_{n}$ are shown in Fig.~\ref{fig:bmag_mupmun}. 
Fig.~\ref{fig:bmag_mupmun} shows that with increasing magnetic field strength, the magnitude of magnetic moment for proton first decreases and then increases, while the magnitude of magnetic moment for neutron first increases and then decreases. Notice the magnetic moment of proton flips the sign when $|eB|\simeq0.062\rm{\,[GeV^2]}$, which is consistent with Fig.~\ref{fig:bmag_mpmn} that when $|eB|\gtrsim0.062\rm{\,[GeV^2]}$, the proton mass increases with the increase of $|eB|$. 

\begin{figure}[htb]
\centering
\includegraphics[scale=0.42]{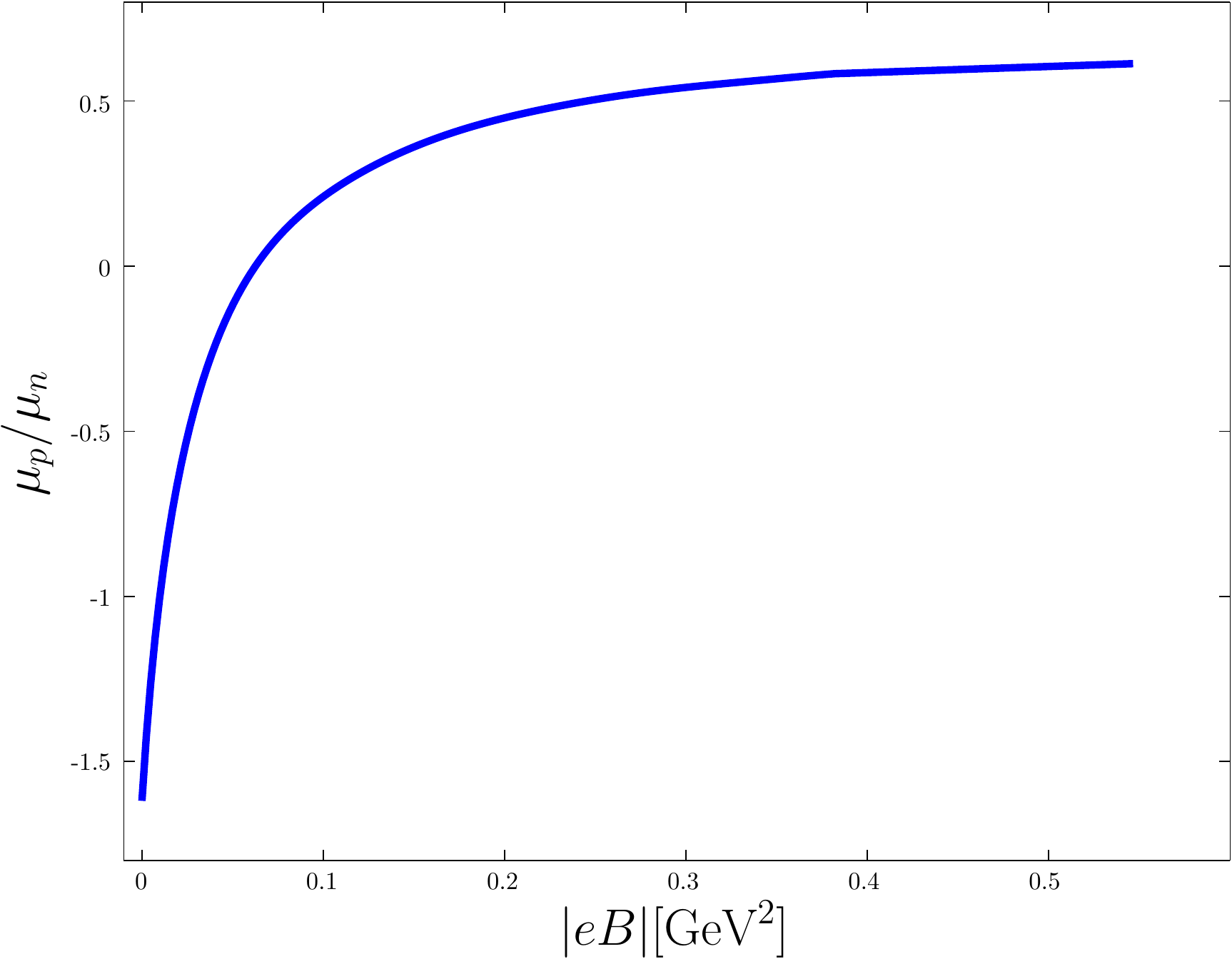} 
\caption{$|e B|$ dependence of $\mu_{p}/\mu_{n}$.
}
\label{fig:bmag_mupOmun}
\end{figure}

The $|eB|$ dependence of $\mu_p/\mu_n$ is shown in Fig.~\ref{fig:bmag_mupOmun}. 
Fig.~\ref{fig:bmag_mupOmun} shows that with the increase of the magnetic field strength, $\mu_{p}/\mu_{n}$ increases. Theoretical analyse of the present model shows that $\mu_{p}/\mu_{n}\to1$ when $|eB|\to \infty$, which agrees with the tendency shown in  Fig.~\ref{fig:bmag_mupOmun}. 

Notice that in Gell-Mann-Nishijima formula \eqref{gmn_formular}, the induced charge from $U(1)$ baryon sector just cancel with the induced charge from the third component of $SU(2)$ iso-vector sector, thus the electric charge density for nucleon state is defined as 
$\rho_E=\frac{1}{2}\rho_{I=0}
+ \<\Psi|I_3|\Psi\>\rho_{I=1}$, 
where $\rho_{I=0}\equiv\left(j_B^0|_{eB\to0}
\right)$, $\rho_{I=1}\equiv\left(\frac{1}{3}\sum_{a=1,2,3}\frac{ \Lambda_a}{\<\Psi|\int dV \Lambda_a|\Psi\>}\right)$ and $\Lambda_a\equiv\frac{\d^2 \hat{\mathscr{L}}}{\d \omega_a^2}$.

The proton root mean square (RMS) electric charge radius and neutron mean square (MS) electric charge radius are defined as $\<r_p^2\>_E^{1/2}\equiv\< p|\int dV X^2 \rho_E |p\>^{1/2}$ and $\<r_n^2\>_E\equiv\< n|\int dV X^2 \rho_E |n\>$, respectively. 
Here $X$ represents $R$, $R_x$ and $R_z$, where $R\equiv\sqrt{x^2+y^2+z^2}$, and $R_x$ and $R_z$ represent the projection of $R$ on the $x$ and $z$ axes, respectively.

\begin{figure}[htb]
\centering
\includegraphics[scale=0.42]{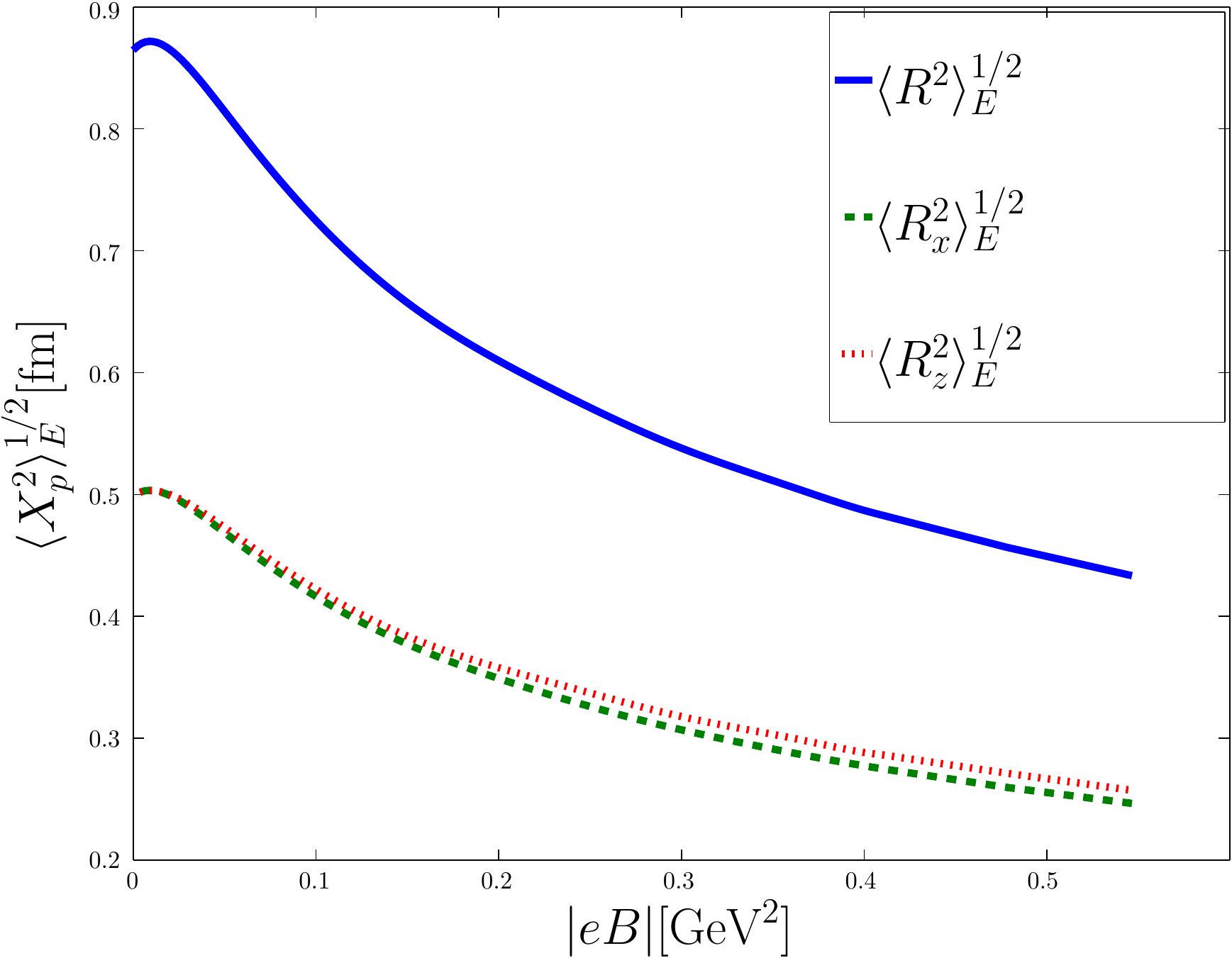} 
\caption{$|e B|$ dependence of the proton RMS electric charge radius $\<r_p^2\>_E^{1/2}$.
}
\label{fig:bmag_rmsp}
\end{figure}

The $|eB|$ dependence of the proton RMS electric charge radii are shown in Fig.~\ref{fig:bmag_rmsp}. 
Fig.~\ref{fig:bmag_rmsp} shows that for proton state: (i) the magnitude of the RMS electric charge radii first increases and then decreases, this tendency is understandable from that: for weak $|eB|$, the proton mass decreases, which causes the proton size to increase; for strong $|eB|$ the freedom of charged meson $\pi^{+,-}$ is restricted in the $x-y$ plane, which causes the proton size to decrease; (ii) the magnitude of $\<R^2_z\>_E^{1/2}$ is slightly larger than $\<R^2_x\>_E^{1/2}$, which is because of that the freedom of charged meson $\pi^{+,-}$ is restricted in the $x-y$ plane, while the neutral meson $\pi^0$ is free to move along $z$ axis, thus, the shape of proton is stretched along $z$ axis.

\begin{figure}[htb]
\centering
\includegraphics[scale=0.42]{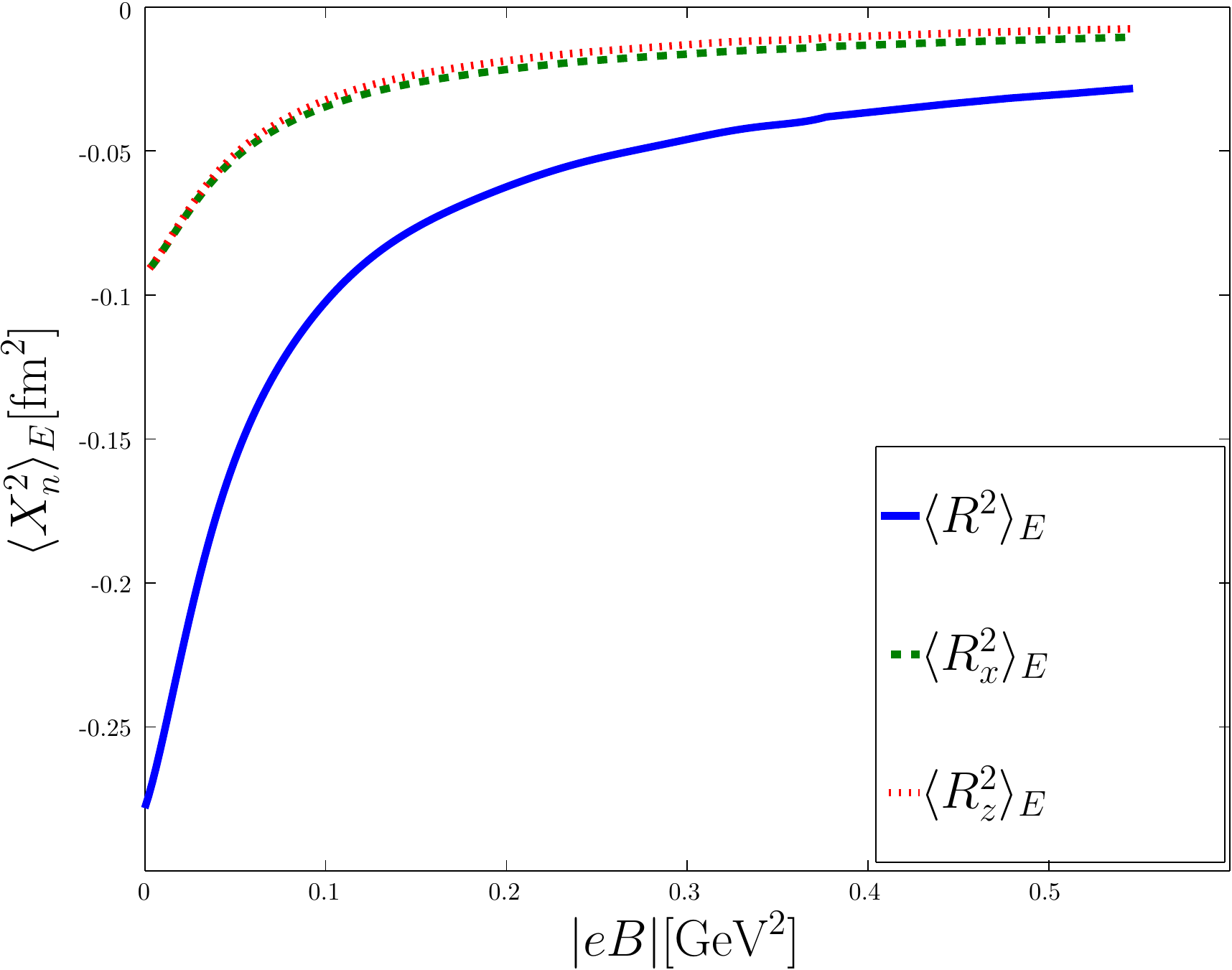} 
\caption{$|e B|$ dependence of the neutron MS electric charge radius $\<r_n^2\>_E$.
}
\label{fig:bmag_msn}
\end{figure}

The $|eB|$ dependence of the neutron MS electric charge radii are shown in Fig.~\ref{fig:bmag_msn}. 
Fig.~\ref{fig:bmag_msn} shows that (i) the neutron MS electric charge radii have a minus sign, this is because of that the distribution of $\rho_{I=1}$ is more apart from the centre  point of the soliton than that of $\rho_{I=0}$; (ii) the magnitude of neutron MS electric charge radii decrease with the increase of $|eB|$, this fact can be understood from that: for all range of $|eB|$, the neutron mass always increases, which causes the neutron size to decrease, i.e., the magnitude of MS electric charge radii will decrease; (iii) the magnitude of $\<R^2_x\>_E$ is slightly larger than $\<R^2_z\>_E$, which is because of that the $\rho_{I=0}$ part is more sensitive with $c_z$ than that of $\rho_{I=1}$ part.

\section{Conclusions and discussions }
In this letter, the properties of proton and neutron in a uniform magnetic field were studied. 

The nucleon states are separated by introduced the $\mathcal{O}(N_C^{-1})$ effects in the semi-classical quantization approach. 

It was first shown that the baryon number and the charge corresponding to the third component of $SU(2)$ iso-vector current are conserved in a uniform magnetic field, respectively. It was found that the induced charge from $U(1)$ baryon sector cancel with the induced charge from the third component of $SU(2)$ iso-vector sector in the Gell-Mann-Nishijima formula.

Next, the $|eB|$ dependence of proton and neutron mass were studied.  
It was found that with the increase of the magnetic field strength, the proton mass first decreases and then increases, while the neutron mass always increases. 
When $|eB|\sim 2.4 m_{\pi}^2$, the proton mass has a minimal point, which decreases about $23{\rm \,[MeV]}$ compared to that in vacuum. 

After that, the proton and neutron magnetic moment were investigated. 
It was found that the magnitude of proton magnetic moment first decreases and then increases, while the magnitude of neutron magnetic moment first increases and then decreases. 
For an extreme weak magnetic field $|eB|\sim 0$, the magnetic moment of proton and neutron are $1.94\,[\mu_N]$ and $-1.21\,[\mu_N]$, respectively, which are consistent with Ref.~\cite{Adkins:1983hy}. 
The ratio of $\mu_p/\mu_n$ is about $-1.60$ when $|eB|\sim 0$, and $0.61$ when $|eB|\sim 28 m_{\pi}^2$. Theoretical analyse of the present model implies $\mu_p/\mu_n\to1$ when $|eB|\to \infty$, which is consistent with the tendency of $\mu_p/\mu_n$.

The RMS electric charge radii of proton and MS electric charge radii of neutron were also investigated. 
It was found that, the proton RMS electric charge radii first increase and then decrease with the increase of the magnetic field. While the magnitude of neutron MS electric charge radii always decrease. 
For an extreme weak magnetic field $|eB|\sim 0$, the proton RMS electric charge radius $\<r_p^2\>_E^{1/2}$ is about $0.865 {\rm \, [fm]}$, which agrees with the experimental result $0.84\sim0.87 {\rm \, [fm]}$; while the neutron MS electric charge radius $\<r_n^2\>_E$ is about $-0.278{\rm \, [fm^2]}$, which is against the experimental result $-0.116{\rm \, [fm^2]}$. 

The present analyse shows that in the core part of the magnetar ($|eB|\sim10^{-2}{\rm \,[ GeV^2}]$), the proton density decreases about $3.4\%$ and the neutron density increases about $15.3\%$ compared to that in vacuum, respectively. Thus, the equation of state in the core part of magnetar is modified. 

In the vacuum of the present analyse, i.e., $|eB|\sim0$, the magnitude of proton and neutron magnetic moment are about $30\%$ and  $36\%$ smaller than the experimental results, respectively. The magnitude of neutron MS electric charge radius is about $1.39$ times larger than the experimental result. 
The inclusion of vector mesons and also scalar mesons might cure these problems~\cite{Meissner:1986js,Meissner:1999pe,Braghin:2003bd,He:2015eua}. 
In the present analysis, the dynamical reaction of magnetic field is neglected, which could change the magnitudes of the  results~\cite{Adam:2014xfa}. These perspectives will be reported elsewhere.

\section{Acknowledgements }
The author thanks Masayasu Harada, Sven Bjarke Gudnason and Muneto Nitta for discussions.


\begin{thebibliography}{99}

\bibitem{Kharzeev:2012ph} 
  D.~E.~Kharzeev, K.~Landsteiner, A.~Schmitt and H.~U.~Yee,
  Lect.\ Notes Phys.\  {\bf 871}, 1 (2013).

\bibitem{Miransky:2015ava} 
  V.~A.~Miransky and I.~A.~Shovkovy,
  Phys.\ Rept.\  {\bf 576}, 1 (2015).

\bibitem{Skokov:2009qp} 
  V.~Skokov, A.~Y.~Illarionov and V.~Toneev,
  Int.\ J.\ Mod.\ Phys.\ A {\bf 24}, 5925 (2009).

\bibitem{Skyrme:1962vh} 
  T.~H.~R.~Skyrme,
  Nucl.\ Phys.\  {\bf 31}, 556 (1962).

\bibitem{He:2015zca} 
  B.~R.~He,
  Phys.\ Rev.\ D {\bf 92}, 111503(R) (2015).

\bibitem{Adkins:1983ya} 
  G.~S.~Adkins, C.~R.~Nappi, and E.~Witten,
  Nucl.\ Phys.\ B {\bf 228}, 552 (1983).


\bibitem{Wess:1971yu} 
  J.~Wess and B.~Zumino,
  Phys.\ Lett.\ B {\bf 37}, 95 (1971).

\bibitem{Witten:1983tw} 
  E.~Witten,
  Nucl.\ Phys.\ B {\bf 223}, 422 (1983).

\bibitem{Holzwarth:1985rb} 
  G.~Holzwarth and B.~Schwesinger,
  Rept.\ Prog.\ Phys.\  {\bf 49}, 825 (1986).

\bibitem{Braaten:1988cc} 
  E.~Braaten and L.~Carson,
  Phys.\ Rev.\ D {\bf 38}, 3525 (1988).

\bibitem{Krusch:2002by} 
  S.~Krusch,
  Annals Phys.\  {\bf 304}, 103 (2003).

\bibitem{Adkins:1983hy} 
  G.~S.~Adkins and C.~R.~Nappi,
  Nucl.\ Phys.\ B {\bf 233}, 109 (1984).


\bibitem{Meissner:1986js} 
  U.~G.~Meissner, N.~Kaiser and W.~Weise,
  Nucl.\ Phys.\ A {\bf 466}, 685 (1987).

\bibitem{Meissner:1999pe} 
  U.~G.~Meissner, A.~Rakhimov and U.~T.~Yakhshiev,
  Phys.\ Lett.\ B {\bf 473}, 200 (2000).

\bibitem{Braghin:2003bd} 
  F.~L.~Braghin and I.~P.~Cavalcante,
  Phys.\ Rev.\ C {\bf 67}, 065207 (2003).

\bibitem{He:2015eua} 
  B.~R.~He, Y.~L.~Ma, and M.~Harada,
  Phys.\ Rev.\ D {\bf 92}, 076007 (2015).

\bibitem{Adam:2014xfa} 
  C.~Adam, T.~Romanczukiewicz, J.~Sanchez-Guillen, and A.~Wereszczynski,
  JHEP {\bf 1411}, 095 (2014).

\end{thebibliography}
\end{document}